\documentclass[preprint2]{aastex}
\usepackage{natbib}
\usepackage{lscape}
\usepackage{graphicx}
\usepackage{morefloats}
\newcommand{\kms}{km s$^{-1}$}

\newcommand{\htco}{H$_2$CO}
\newcommand{\hdco}{HDCO}
\newcommand{\trani}{$2_{1,2}\rightarrow1_{1,1}$}
\newcommand{\tranii}{$2_{0,2}\rightarrow1_{0,1}$}
\newcommand{\traniii}{$2_{1,1}\rightarrow1_{1,0}$}
\newcommand{\jk}{$J_{K_{\rm a},K_{\rm c}}$}

\shorttitle{Deuterium fractionation of protostars}
\shortauthors{Kang et al.}
\slugcomment{To appear in ApJ}

\begin{document}

\title{Measurement of [HDCO]/[H$_2$CO] Ratios
       in the Envelopes of Extremely Cold Protostars in Orion}

\author{\sc Miju Kang$^{1,2,3}$,  
            Minho Choi$^{1}$,
            Amelia M. Stutz$^{4}$,
            and Ken'ichi Tatematsu$^{5,6}$}

\affil{$^1$ Korea Astronomy and Space Science Institute,
            776 Daedeokdaero, Yuseong, Daejeon 305-348, Republic of Korea;
            mjkang@kasi.re.kr \\
		 $^2$ Korea University of Science and Technology,
		  		Daejeon 305-350, Republic of Korea \\
		 $^3$ Max-Planck-Institut f\"{u}r Radioastronomie,
		  		Auf dem H\"{u}gel 69, 53121 Bonn, Germany \\
		 $^4$ Max Planck Institute for Astronomy, K\"{o}nigstuhl 17,
		  		D-69117 Heidelberg, Germany \\
		 $^5$ National Astronomical Observatory of Japan, 2-21-1 Osawa,
		  		Mitaka, Tokyo 181-8588, Japan \\
		 $^6$ Department of Astronomical Science,
		      The Graduate University for Advanced Studies (SOKENDAI),
		      2-21-1 Osawa, Mitaka, Tokyo 181-8588, Japan}

\begin{abstract}

We present observations of \hdco\ and \htco\ emission toward a sample
of 15 Class 0 protostars in the Orion A and B clouds. Of these, eleven
protostars are {\it Herschel}-identified PACS Bright Red
Sources (PBRS) and four are previously identified protostars.  Our
observations revealed the chemical properties of the PBRS envelope
for the first time. The column densities of \hdco\ and \htco\ are
derived from single dish observations at an angular resolution of
$\sim$20$\arcsec$ ($\sim$8400~AU). The degree of deuteration in
\htco\ ([HDCO]/[H$_2$CO]) was estimated to range from 0.03 to 0.31. The
deuterium fractionation of most PBRS (70\%) is similar to that of the
non-PBRS sources. Three PBRS (30\%) exhibit
high deuterium fractionation, larger than 0.15. The large variation
of the deuterium fractionation of \htco\ in the whole PBRS sample may
reflect the diversity in the initial conditions of star forming
cores. There is no clear correlation between the [HDCO]/[H$_2$CO]
ratio and the evolutionary sequence of protostars. 

\end{abstract}

\keywords{astrochemistry --- ISM: abundances --- ISM: molecules
          --- stars: formation}

\section{Introduction}

Observing deuterated species is useful for probing the physical and
chemical conditions of star-forming regions \citep{Roberts:2000un,
  Caselli:2002eg, Roueff:2003ig, Crapsi:2005kp, Tatematsu:2010jr,
  Fontani:2011ie, Caselli:2012bf, Sakai:2012eh}.
The deuterium to hydrogen (D/H) ratio in the local interstellar medium
is low, about $1.5 \times 10^{-5}$,
which is the D/H elemental abundance as predicted by the nucleosynthesis
\citep{Linsky:2003kn}.
By contrast, higher molecular D/H ratios have been observed
from objects associated with star formation,
such as molecular clouds
\citep{Loren:1985gm, Turner:2001he, Parise:2009co, Bergman:2011fh},
pre-stellar cores \citep{Bacmann:2003kv, Crapsi:2005kp, Fontani:2008dx},
and protostellar cores
\citep{Ceccarelli:1998vd, Loinard:2002kt, Emprechtinger:2009bp}.
Many observations and chemical models showed that deuterium-bearing
molecules are enhanced during the cold prestellar core phase and
released into the gas when a heating event, such as the formation
of a protostar or the passage of a shock, evaporates the ice
\citep{Tielens:1983te, Rodgers:1996vw, Ceccarelli:2001gk,
  Bacmann:2003kv, Maret:2004io, Cazaux:2011dt, Taquet:2012gv,
  Awad:2014gk, Fontani:2014jh}.

Formaldehyde (\htco) is one of the key species in the synthesis of
more complex organic molecules. The \htco\ lines are ubiquitous in
the interstellar medium and used to infer temperature and density
of gas \citep{Mangum:1993jp, Ceccarelli:2003ir}. There have been
debates on the dominant formation mechanism of deuterated \htco:
whether singly or doubly deuterated species (HDCO or D$_2$CO) of
formaldehyde form via active grain surface chemistry \citep{Tielens:1983te,
Turner:1990dg, Ceccarelli:2001gk, Loinard:2001im, Fontani:2015kr}
or gas-phase reactions \citep{Langer:1979fn, Roberts:2000wt,
Parise:2009co}.
\cite{Codella:2012he} investigated
deuteration in the protostellar outflow shock of L1157-B1.  They
concluded that the measured deuterium fractionation provides a
fossil record of the gas before it was shocked by the jet driven
by the protostar. They noted that the deuteration ratios derived
from the outer part of the L1157 dense envelope (L1157-B1) are
smaller than the ratios derived from the IRAS 16293-2422 spectra
which is close to the protostar, where the density is expected to
be higher.  Recently, high resolution observations with a
$\sim$2$\arcsec$ resolution toward L1157-B1 showed that the emission
of HDCO was detected mostly from the region of interface between
the ambient material and the shock driven by active outflow from a
low-mass Class 0 protostar L1157-mm \citep{Fontani:2014jh}. The
deuterated fraction [HDCO]/[H$_2$CO] of $\sim$0.1 was measured in
the HDCO emitting region. These observations confirmed that deuterated
molecules are formed on the grain surface and released into gas
phase by the shocks.

The Orion molecular cloud complex is one of the nearest active
star-forming regions at a distance of $\sim$420 pc
\citep{Hirota:2007wm,Kim:2008gs,Menten:2007ew}.  \cite{Stutz:2013ka}
discovered and characterized the PACS Bright Red Sources (PBRS) in
Orion. Selected to have extremely red 70 \micron\ to 24 \micron\
colors (log($\lambda F_{\lambda}(70\, \micron)/\lambda F_{\lambda}
(24\, \micron)) > 1.65$), the PBRS have $T_{\rm bol} < 45$ K and
$L_{\rm smm}/L_{\rm bol} > 0.6\%$. They are colder than the typical
Class 0 protostars in Orion. \cite{Stutz:2013ka} proposed that the
PBRS may be the youngest protostars identified and characterized
to date, with relatively high envelope densities. The PBRS are,
therefore, ideal objects for investigating the earliest stage of
protostellar evolution.  The aim of this work is to investigate
whether the youngest Class 0 objects, PBRS, are different from
non-PBRS Class 0 protostars, by comparing the deuterated levels of
\htco.

In this paper, we report the observations of the selected protostars
in the \htco\ \jk\ $=$ \trani\ transition and the \hdco\ \jk\ $=$
\tranii\ and \traniii\ transitions with two antennas of the Korean
Very Long Baseline Interferometry Network (KVN)
used in the single-dish mode.
We describe source selection and details of our KVN
observations in Section 2.  The main observational results are
presented in Section 3. In Section 4 we derive the physical properties.
In Section 5 we discuss the deuterium fractionation of protostars.
A summary is given in Section 6.

\section{Source Selection and Observations}

\begin{deluxetable}{ccrrccc}
\tabletypesize{\small}
\tablecaption{List of Observed Sources}
\tablewidth{0pt}
\tablehead{
\multicolumn{2}{c}{Source\tablenotemark{a}}
& \colhead{R.A.}
& \colhead{Declination}
& \colhead{$T_{\rm bol}$\tablenotemark{b}}
& \colhead{$L_{\rm bol}$\tablenotemark{b}}
& \colhead{Field}
\\
\colhead{HOPS ID}
& \colhead{PBRS ID}
& \colhead{(h:m:s)}
& \colhead{($\degr : \arcmin : \arcsec$)}
& \colhead{(K)}
& \colhead{($L_{\odot}$)}
&
}
\startdata
\multicolumn{7}{c}{PBRS} \\
\hline
398  & 119019       &   05:40:58.4 & $-$08:05:36.1& 34 & ~1.6 & L1641 \\
169  &              &   05:36:36.0 & $-$06:38:54.0& 35 & ~4.5 & L1641 \\
400  & 082005       &   05:41:29.4 & $-$02:21:17.1& 29 & ~1.0 & NGC 2024 \\
403  & 090003       &   05:42:45.2 & $-$01:16:14.2& 36 & ~2.7 & NGC 2024 \\
358  &              &   05:46:07.2 & $-$00:13:30.9& 44 & 30.6 & NGC 2068 \\
373  &              &   05:46:30.7 & $-$00:02:36.8& 36 & ~5.2 & NGC 2068 \\
404  & 093005       &   05:46:27.7 & $-$00:00:53.8& 31 & ~1.7 & NGC 2068 \\
359  &              &   05:47:24.8 &    00:20:58.2& 39 & 12.6 & NGC 2068 \\
341  &              &   05:47:00.9 &    00:26:20.8& 36 & ~3.6 & NGC 2068 \\
405  & 097002       &   05:48:07.7 &    00:33:50.8& 33 & ~1.1 & NGC 2068 \\
354  &              &   05:54:24.1 &    01:44:20.2& 37 & ~7.5 & L1622 \\
\hline
\multicolumn{7}{c}{Non-PBRS\tablenotemark{c}} \\
\hline
164 &               &   05:37:00.5 & $-$06:37:10.5  & 50 & ~0.6 & L1641 \\
297 &               &   05:41:23.3 & $-$02:17:35.8  & 46 & ~1.0 & NGC 2024 \\
321 &               &   05:46:33.2 &    00:00:02.2  & 74 & ~4.6 & NGC 2068 \\
360 &               &   05:47:27.1 &    00:20:33.1  & 36 & ~2.2 & NGC 2068 \\
\enddata
\tablenotetext{a}{In case of having two source IDs,
                  we use the PBRS ID in the text.}
\tablenotetext{b}{Uncertainties in $T_{\rm bol}$ and $L_{\rm bol}$
                  are $\sim$15\% \citep{Stutz:2013ka}.}
\tablenotetext{c}{$T_{\rm bol}$ and $L_{\rm bol}$ of non-PBRS
                  are from A. Stutz (private communication)
                  and Furlan et al. (2015, in preparation).}
\label{table_sources}
\end{deluxetable}

\begin{deluxetable}{rccccc}
\tabletypesize{\small}
\tablecaption{Observed Transitions and Telescope Parameters}
\tablewidth{0pt}
\tablehead{
\colhead{Transition} & \colhead{Rest frequency} & \colhead{$A_{\rm ul}$} & \colhead{$E_{\rm l}$} & \colhead{$\eta_{\rm mb}$}    & \colhead{$T_{\rm sys}$\tablenotemark{a}}\\
                     & \colhead{(GHz)}          & \colhead{(s$^{-1}$)}   & \colhead{(K)}         &\colhead{KUS, KYS}            & \colhead{(K)}              }
\startdata
\hdco\ \tranii  & 128.812865 & $5.4 \times 10^{-5}$  & ~3.1  & 0.30, 0.33 &  240 \\
\hdco\ \traniii & 134.284909 & $4.6 \times 10^{-5}$  & 11.2  & 0.28, 0.31 &  230 \\
\htco\ \trani   & 140.839520 & $5.3 \times 10^{-5}$  & 15.2  & 0.25, 0.28 &  300 \\
\enddata

\label{table_obs_lines}
\tablenotetext{a}{Mean system temperature}
\end{deluxetable}

The {\it Herschel} Orion Protostar Survey (HOPS) is a {\it Herschel}
Open Time Key Program targeting about 300 {\it Spitzer}-identified
protostars with PACS 70 \micron\ and 160 \micron\ photometry and a
subset of 30 protostars with PACS spectroscopy \citep{Fischer:2010bl,
Manoj:2013ie, Stutz:2013ka}. \cite{Stutz:2013ka} found 18 PBRS in
the Orion molecular clouds using {\it Herschel} scan-map observations
toward a subset of the HOPS fields. We selected 11 isolated PBRS that
are situated away from other protostars to avoid confusion in the
KVN beam ($\sim$20\arcsec).
For comparison, we added 4 non-PBRS HOPS targets,
located in the vicinity of the PBRS targets, to the sample.
The non-PBRS targets were selected
based on their bolometric temperature and distance to the PBRS targets.
They are located in the L1641, NGC 2024, and NGC 2068 fields of Orion,
where most of the PBRS targets are also located.
The number of non-PBRS targets was limited by the available observing time.
Both PBRS and non-PBRS targets are taken from the HOPS sample.
The coordinates and
physical parameters of the selected sources are listed in Table
\ref{table_sources}.

Observations were made with the KVN 21 m antennas
in the single-dish telescope mode during the 2013 observing season,
with the KVN Yonsei antenna at Seoul in 2013 September--October
and the KVN Ulsan antenna in 2013 November--December.
The telescope pointing was checked once every two hours
by observing Orion IRc2 \citep{Baudry:1995tz}
in the SiO $v=1$ $J=1\rightarrow0$ maser line
and was good within $\sim$5\arcsec.
The antenna temperature was calibrated
by the standard chopper-wheel method.

We used the spectrometer with a bandwidth of 64 MHz and 4096 channels,
which gives a spectral channel width of 15.625 kHz. Table
\ref{table_obs_lines} lists the parameters of the observed lines.
The spectrometer setting gives a velocity channel width
of 0.036 \kms, 0.035 \kms, and 0.033 \kms\
for \hdco\ \tranii, \hdco\ \traniii, and \htco\ \trani, respectively.
For all lines, Hanning smoothing
was applied three times. The half power beam width is about 20\arcsec.
Quantization correction factor (1.25) and sideband separation
efficiency (0.8 for the KVN Ulsan 129 GHz band) were applied to the
KVN raw data.
The data acquired in the antenna temperature scale, $T_A^*$,
were converted to the main beam temperatures scale,
$T_{\rm mb} = T_A^* / \eta_{\rm mb}$,
using the main beam efficiencies ($\eta_{\rm mb}$)
in Table \ref{table_obs_lines}.
The data were processed with the
GILDAS/CLASS software from Institute de RadioAstronomie Millim\'etrique
(http://www.iram.fr/IRAMFR/GILDAS).

\section{Results}
\label{sec:results}


\begin{figure*}[!t]
\epsscale{0.6}

\plotone{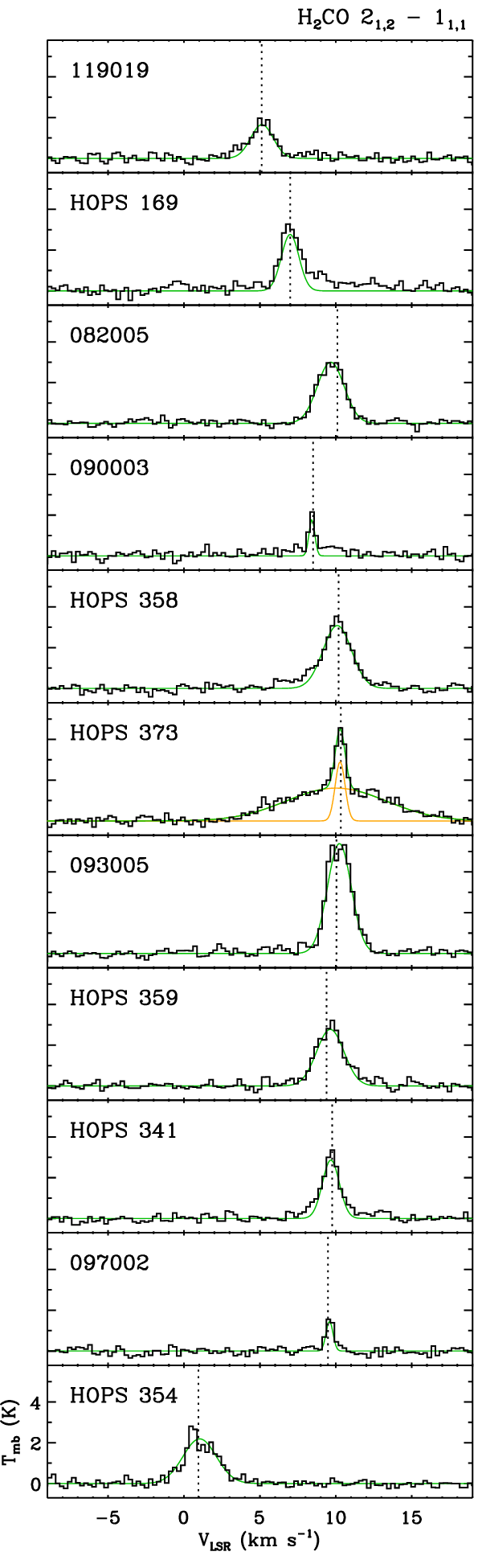} 	   
\plotone{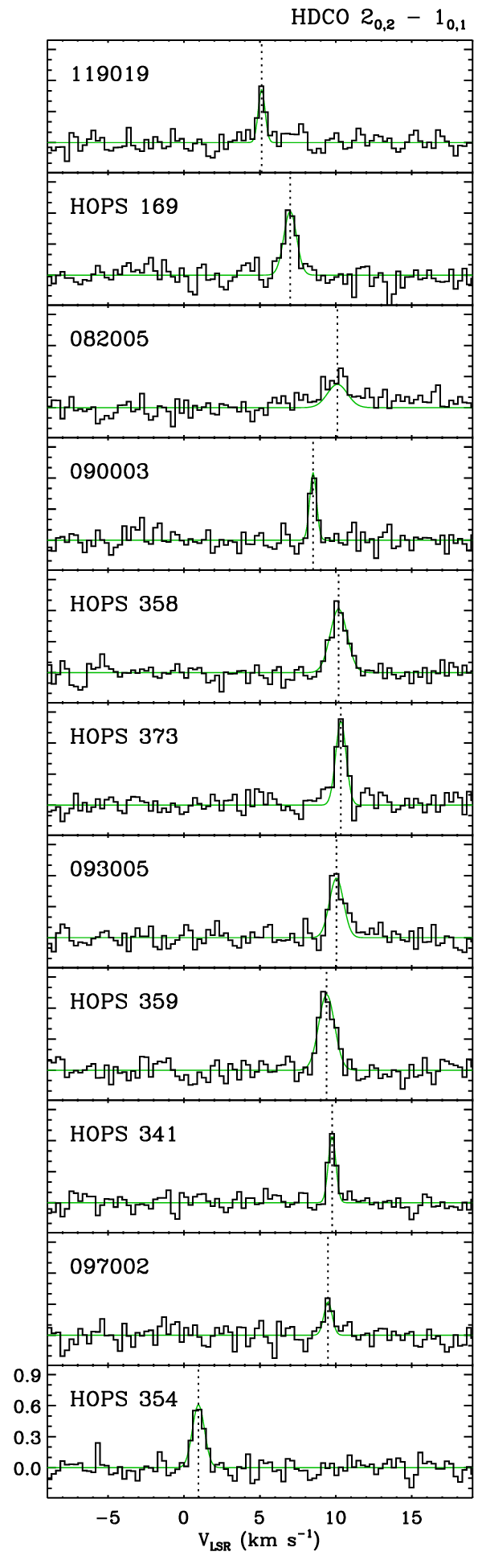}	
\plotone{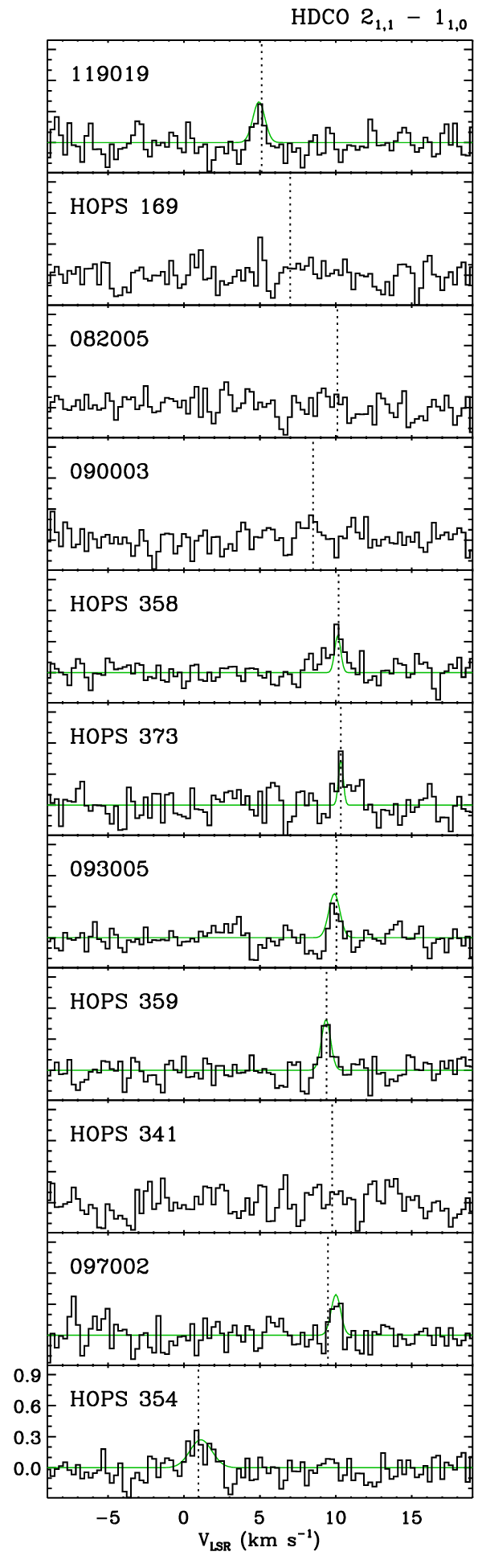}	

\caption{Spectra of the \htco\ \trani\ transition,
	 the \hdco\ \tranii\ and \traniii\ transitions observed
	 toward PBRS targets.  Vertical dotted lines show the
	 centroid velocities of the \hdco\ \tranii\ spectra.  The
	 green spectra are Gaussian fits.  The orange spectra of
	 HOPS 373 show individual components of the two-component
	 Gaussian fit.}

\label{fig_spectra_pbrs}

\end{figure*}
 

\begin{figure*}[!t]
\epsscale{0.6}

\plotone{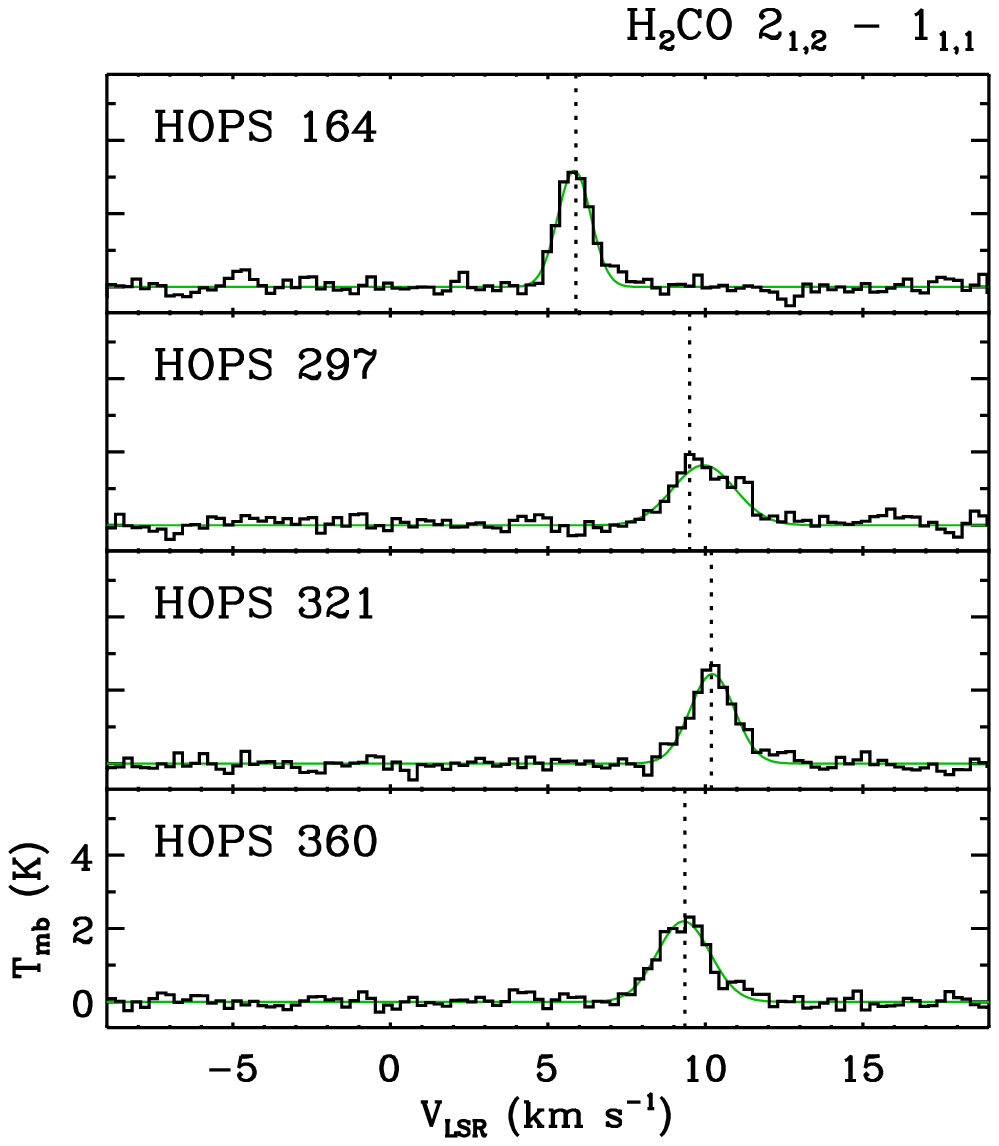}
\plotone{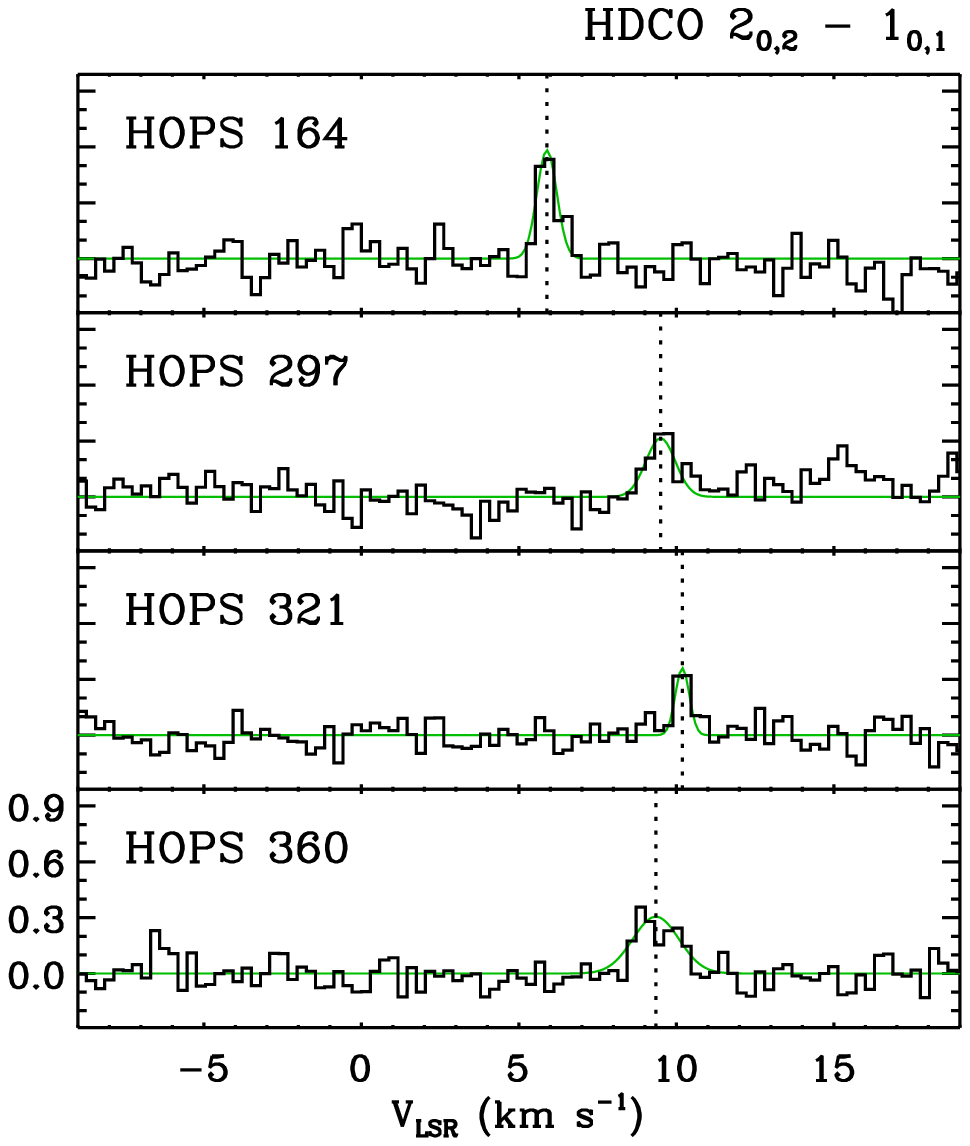}
\plotone{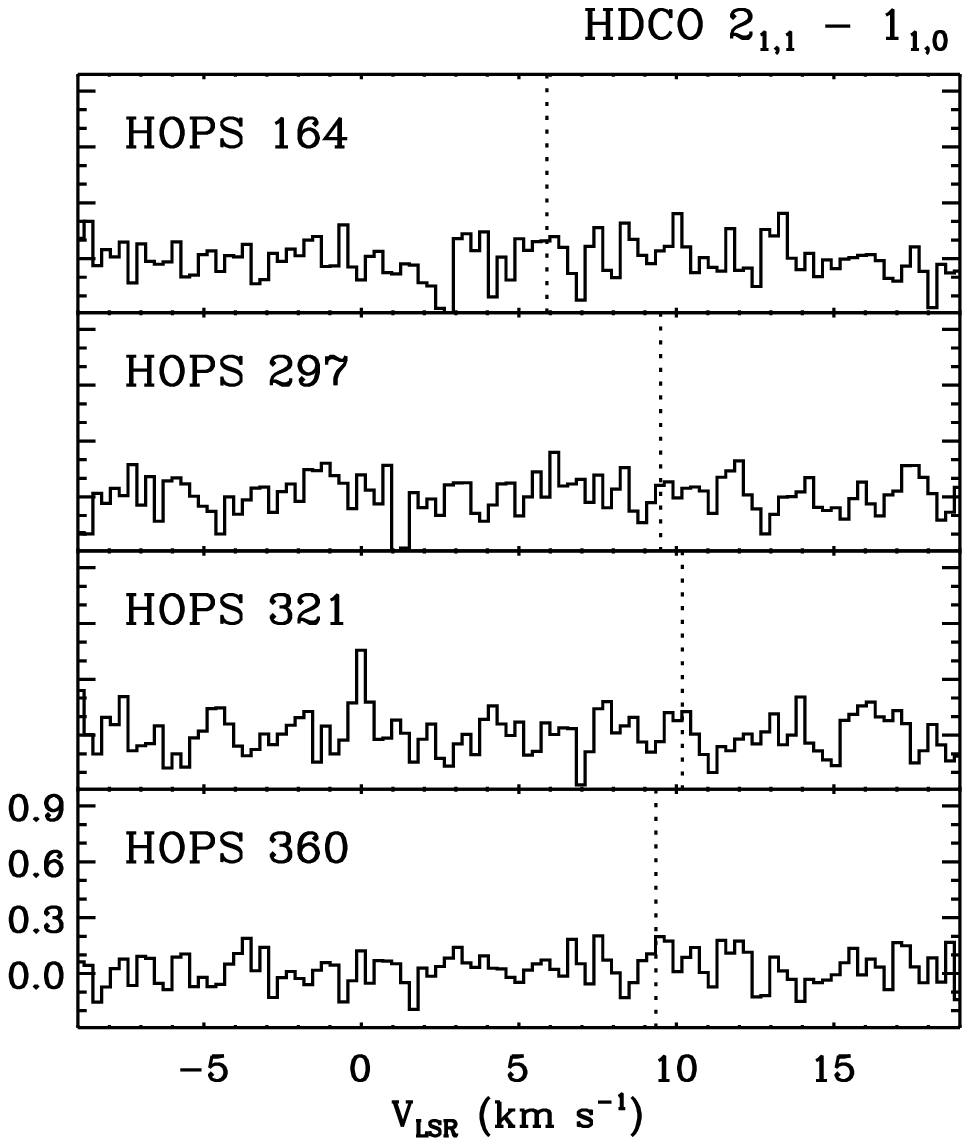}

\caption{Same as Fig. 1 but for non-PBRS targets.}

\label{fig_spectra_hops}

\end{figure*}

\begin{deluxetable}{crrrcrrrcccc}
\renewcommand{\arraystretch}{1.1}
\renewcommand{\tabcolsep}{0.8mm}
\tabletypesize{\scriptsize}
\tablecaption{Parameters of the Observed Spectra}
\tablewidth{0pt}
\tablehead{
\colhead{Source}
& \multicolumn{3}{c}{\htco\ \trani} &
& \multicolumn{3}{c}{\hdco\ \tranii} &
& \multicolumn{3}{c}{\hdco\ \traniii}
\\
\cline{2-4} \cline{6-8}  \cline{10-12}
& \colhead{$\int T_{\rm mb} {\rm d}v$} & \colhead{$v_{c}$} & \colhead{$\Delta v$} &
& \colhead{$\int T_{\rm mb} {\rm d}v$} & \colhead{$v_{c}$} & \colhead{$\Delta v$} &
& \colhead{$\int T_{\rm mb} {\rm d}v$} & \colhead{$v_{c}$} & \colhead{$\Delta v$}
\\
& \colhead{(K km s$^{-1}$)} & \colhead{(km s$^{-1}$)} & \colhead{(km s$^{-1}$)}&
& \colhead{(K km s$^{-1}$)} & \colhead{(km s$^{-1}$)} & \colhead{(km s$^{-1}$)}&
& \colhead{(K km s$^{-1}$)} & \colhead{(km s$^{-1}$)} & \colhead{(km s$^{-1}$)}
}
\startdata
119019                     &  3.07 $\pm$ 0.16 &   5.13 $\pm$ 0.04 & 1.74 $\pm$ 0.11 &  &0.29 $\pm$ 0.03 &  5.11 $\pm$ 0.03 & 0.53 $\pm$ 0.06 &  &0.39 $\pm$ 0.06 & ~4.93 $\pm$ 0.07  & 0.93 $\pm$ 0.17 \\
HOPS 169                   &  4.02 $\pm$ 0.16 &   6.99 $\pm$ 0.03 & 1.36 $\pm$ 0.06 &  &0.65 $\pm$ 0.07 &  6.99 $\pm$ 0.05 & 0.99 $\pm$ 0.13 &  &\nodata         &  \nodata          & \nodata         \\
082005                     &  6.66 $\pm$ 0.10 &   9.66 $\pm$ 0.02 & 2.08 $\pm$ 0.04 &  &0.37 $\pm$ 0.06 & 10.10 $\pm$ 0.12 & 1.54 $\pm$ 0.27 &  &\nodata         &  \nodata          & \nodata         \\
090003                     &  0.86 $\pm$ 0.05 &   8.42 $\pm$ 0.01 & 0.45 $\pm$ 0.02 &  &0.37 $\pm$ 0.04 &  8.50 $\pm$ 0.03 & 0.53 $\pm$ 0.07 &  &\nodata         &  \nodata          & \nodata         \\
HOPS 358                   &  7.27 $\pm$ 0.20 &  10.09 $\pm$ 0.03 & 2.21 $\pm$ 0.08 &  &0.86 $\pm$ 0.04 & 10.18 $\pm$ 0.03 & 1.31 $\pm$ 0.07 &  &0.18 $\pm$ 0.05 & 10.12 $\pm$ 0.06  & 0.47 $\pm$ 0.18 \\
HOPS 373\tablenotemark{a}  &  2.36 $\pm$ 0.13 &  10.28 $\pm$ 0.02 & 0.76 $\pm$ 0.04 &  &0.70 $\pm$ 0.06 & 10.33 $\pm$ 0.03 & 0.81 $\pm$ 0.08 &  &0.17 $\pm$ 0.07 & 10.33 $\pm$ 0.09  & 0.36 $\pm$ 0.12 \\
093005                     & 10.58 $\pm$ 0.15 &  10.23 $\pm$ 0.01 & 1.84 $\pm$ 0.03 &  &0.67 $\pm$ 0.05 & 10.04 $\pm$ 0.05 & 1.09 $\pm$ 0.11 &  &0.39 $\pm$ 0.05 & ~9.90 $\pm$ 0.06  & 0.86 $\pm$ 0.13 \\
HOPS 359                   &  6.28 $\pm$ 0.27 &   9.63 $\pm$ 0.04 & 2.12 $\pm$ 0.11 &  &0.96 $\pm$ 0.06 &  9.39 $\pm$ 0.04 & 1.23 $\pm$ 0.08 &  &0.37 $\pm$ 0.09 & ~9.35 $\pm$ 0.07  & 0.71 $\pm$ 0.22 \\
HOPS 341                   &  4.01 $\pm$ 0.12 &   9.67 $\pm$ 0.02 & 1.31 $\pm$ 0.05 &  &0.40 $\pm$ 0.03 &  9.75 $\pm$ 0.02 & 0.58 $\pm$ 0.05 &  &\nodata         &  \nodata          & \nodata         \\
097002                     &  0.83 $\pm$ 0.04 &   9.58 $\pm$ 0.01 & 0.53 $\pm$ 0.03 &  &0.22 $\pm$ 0.04 &  9.48 $\pm$ 0.05 & 0.64 $\pm$ 0.14 &  &0.31 $\pm$ 0.06 & 10.00 $\pm$ 0.08  & 0.74 $\pm$ 0.13 \\
HOPS 354                   &  6.03 $\pm$ 0.16 &   1.03 $\pm$ 0.03 & 2.58 $\pm$ 0.08 &  &0.61 $\pm$ 0.06 &  0.95 $\pm$ 0.04 & 0.94 $\pm$ 0.10 &  &0.49 $\pm$ 0.09 & ~1.11 $\pm$ 0.18  & 1.70 $\pm$ 0.30 \\
\hline
HOPS 164                   &  4.20 $\pm$ 0.13 &   5.84 $\pm$ 0.02 & 1.24 $\pm$ 0.04 &  &0.46 $\pm$ 0.05 &  5.89 $\pm$ 0.04 & 0.74 $\pm$ 0.11 &  &\nodata         &  \nodata           & \nodata        \\
HOPS 297                   &  4.07 $\pm$ 0.19 &   9.95 $\pm$ 0.05 & 2.33 $\pm$ 0.12 &  &0.37 $\pm$ 0.05 &  9.50 $\pm$ 0.07 & 1.10 $\pm$ 0.20 &  &\nodata         &  \nodata           & \nodata        \\
HOPS 321                   &  4.30 $\pm$ 0.16 &  10.22 $\pm$ 0.03 & 1.65 $\pm$ 0.08 &  &0.20 $\pm$ 0.03 & 10.19 $\pm$ 0.04 & 0.52 $\pm$ 0.12 &  &\nodata         &  \nodata           & \nodata        \\
HOPS 360                   &  4.64 $\pm$ 0.14 &   9.31 $\pm$ 0.03 & 1.98 $\pm$ 0.07 &  &0.56 $\pm$ 0.07 &  9.35 $\pm$ 0.11 & 1.73 $\pm$ 0.24 &  &\nodata         &  \nodata           & \nodata
\enddata
\tablecomments{The calibration uncertainty of 10\%
               is not included \citep{Lee:2011dn}.}
\tablenotetext{a}{HOPS 373 shows an outflow feature.
                  This table lists the parameters for the narrow component.
                  The integrated intensity, centroid velocity,
                  and line width for the line wing component are
                  13.31 $\pm$ 0.34 K \kms, 10.02 $\pm$ 0.09 \kms,
                  and 7.75 $\pm$ 0.22 \kms, respectively.}

\label{table_obs_results}
\end{deluxetable}

The observed \htco\ \trani, \hdco\ \tranii, and \traniii\ line
spectra are shown in Figure \ref{fig_spectra_pbrs} for PBRS and
Figure \ref{fig_spectra_hops} for non-PBRS. The results of the
observations are summarized in Table \ref{table_obs_results},
including integrated intensities, centroid velocities, and line widths
derived by Gaussian fits to the spectra.

\htco\ \trani\ emission was detected toward all targets.  The line
widths of all sources are smaller than 3 \kms.
Two PBRS, 090003 and 097002, have quite small line widths, about 0.5 \kms.
Several sources with relatively large line widths show flat top spectra,
indicating either large optical depths or overlap of multiple components.
For example, the \htco\ spectrum of 093005 shows double peaks.
Since the velocity of the central dip
corresponds to the peak velocity of the \hdco\ \tranii\ spectrum,
we prefer the interpretation of a large optical depth.
HOPS 373 drives a CO outflow \citep{Gibb:2000jf},
and the \htco\ spectrum shows broad line wings extending to large velocities.
For HOPS 373, we applied a two-component Gaussian fit
and used the parameters derived from the narrow component,
which allows us to focus on the emission
arising from the envelope of the protostar.

\hdco\ \tranii\ emission was detected toward all sources. Line
widths of the \hdco\ \tranii\ transition are narrower than those
of the \htco\ \trani\ transition. The narrow lines emitted by the
deuterated formaldehyde suggest that the emission is optically thin
and dominated by the cold outer envelopes of the protostars.  \hdco\
\traniii\ emission was detected toward seven PBRS. The qualities
of the HDCO \traniii\ spectra toward 119019, 097002, and HOPS 354
are relatively poor. Spectral baselines (line free channels) of
these sources seem to be worse than the other cases. The HDCO
\traniii\ emission of 097002 may be a non-detection, because the
line is placed at a peak velocity slightly different from that of
the other transitions detected toward this source.

\section{Analysis}

\subsection{H$_2$CO column density}

We calculate the column densities of \htco.
Assuming that the line is optically thin and the source
size is similar to the main beam size, the column density in a lower
state (defined as energy level $u\rightarrow\ l = (J+1)_{K_{\rm
a},K_{\rm c}+1} \rightarrow\ J_{K_{\rm a},K_{\rm c}}$) can be
obtained by
\begin{equation}
N_l = 93.5 \frac{g_l}{g_u} \frac{\nu^3}{A_{ul}}
      \frac{1}{T_{\rm ex}[1-{\rm exp}(-h\nu/kT_{\rm ex})]}
      \int_{}T_{\rm mb} \,\mathrm{d}v,
\end{equation}
where $h$ is the Planck's constant, $k$ the Boltzmann's constant,
$\nu$ the frequency of the transition in GHz,
$T_{\rm ex}$ the excitation temperature, $g$ the statistical weight,
$A_{ul}$ the Einstein coefficient in s$^{-1}$,
$\int_{}T_{\rm mb} \,\mathrm{d}v$ the integrated line intensity,
and $v$ the velocity in km s$^{-1}$ \citep{Rohlfs:1999}.
The column density is in cm$^{-2}$,
and the numerical factor in the equation
corresponds to 8$\pi$/$c^3$ and a conversion factor,
when it is expressed in the units given above.
All parameters for molecules were taken from the JPL databases
\citep{Pickett:1998jh}.
Some sources may have moderate optical depths as suggested in Section \ref{sec:results},
and they are discussed in Section \ref{sec:tau}. 

The total column density of ortho-\htco, $N_{\rm ortho}$, is
related to the column density, ${N_l} = N(J_{K_{\rm a},K_{\rm c}})$, in the
lower state $J_{K_{\rm a},K_{\rm c}}$ by
\begin{equation} 
\frac{N_{\rm ortho}} {N(J_{K_{\rm a},K_{\rm c}})} = \frac{Z}{g({J_{K_{\rm
a},K_{\rm c}}})}{{\rm exp}\left(\frac{E(J_{K_{\rm a},K_{\rm
c}})}{kT_{\rm ex}}\right)}, 
\end{equation}
where $g({J_{K_{\rm a},K_{\rm c}}})$
is the statistical weight ($3(2J+1)$ for ortho-H$_2$CO),
$E(J_{K_{\rm a}K_{\rm c}})$ the energy of a $J_{K_{\rm a}K_{\rm c}}$ level,
and $Z$ the partition function.
If the molecules follow the Boltzmann distribution of a
single $T_{\rm ex}$, the partition function is
\begin{equation}
Z = \sum_{} z = \sum_{}g({J_{K_{\rm a},K_{\rm c}}}) {\rm exp}\left(-\frac{E(J_{K_{\rm a}K_{\rm
c}})}{kT_{\rm ex}}\right).
\end{equation}
For ortho-\htco, $K_{\rm a}$ can only be odd.
The total \htco\ column densities are calculated assuming an excitation
temperature of 10 K and the statistical ortho to para ratio of 3:1
\citep{Minh:1995ty, Guzman:2013ib}. The resulting column densities
are given in Table \ref{table_cd_ratio}.
The column densities are not very sensitive
to the assumed excitation temperature,
when $T_{\rm ex} \gtrsim$ 10 K \citep{Roberts:2002ck}.
Using an excitation temperature of 30 K or 50 K
changes the inferred column densities by a factor of about 2.2 or 3.7, respectively.
For example, if we calculate column densities of \htco\
adopting $T_{\rm ex} = 30$ K for 090003 and HOPS 341,
their column densities become
$7.1 \times 10^{12}$ and $33.0 \times 10^{12}$ cm$^{-2}$, respectively.
Indeed, we estimate $T_{\rm ex} \approx$ 14 K from
those sources with detections of both HDCO transitions
(Section \ref{sec:Tex}).
Therefore, the assumption of $T_{\rm ex}$ = 10 K is reasonable.

\begin{deluxetable}{cccc}
\tabletypesize{\small}
\tablecaption{Column Densities}
\tablewidth{0pt}
\tablehead
{\colhead{Source} & \colhead{$N$(H$_2$CO)}             & \colhead{$N$(HDCO)}                & \colhead{[HDCO]/[H$_2$CO]} \\
                  & \colhead{($10^{12} {\rm cm}^{-2}$)} & \colhead{($10^{12} {\rm cm}^{-2}$)} & \colhead{}       }
\startdata
    119019        &  11.7 $\pm$ 1.3 & 0.79 $\pm$ 0.11 & 0.07 $\pm$  0.01\\
    HOPS 169      &  15.3 $\pm$ 1.6 & 1.76 $\pm$ 0.26 & 0.12 $\pm$  0.02\\
    082005        &  25.3 $\pm$ 2.6 & 1.00 $\pm$ 0.19 & 0.04 $\pm$  0.01\\
    090003        &  ~3.3 $\pm$ 0.4 & 1.00 $\pm$ 0.15 & 0.31 $\pm$  0.06\\
    HOPS 358      &  27.6 $\pm$ 2.9 & 2.33 $\pm$ 0.26 & 0.08 $\pm$  0.01\\
    HOPS 373      &  ~9.0 $\pm$ 1.0 & 1.90 $\pm$ 0.25 & 0.21 $\pm$  0.04\\
    093005        &  40.2 $\pm$ 4.1 & 1.82 $\pm$ 0.23 & 0.05 $\pm$  0.01\\
    HOPS 359      &  23.9 $\pm$ 2.6 & 2.61 $\pm$ 0.31 & 0.11 $\pm$  0.02\\
    HOPS 341      &  15.3 $\pm$ 1.6 & 1.09 $\pm$ 0.14 & 0.07 $\pm$  0.01\\
    097002        &  ~3.2 $\pm$ 0.4 & 0.60 $\pm$ 0.12 & 0.19 $\pm$  0.04\\
    HOPS 354      &  23.0 $\pm$ 2.4 & 1.65 $\pm$ 0.23 & 0.07 $\pm$  0.01\\
\hline
    HOPS 164      &  16.0 $\pm$ 1.7 & 1.25 $\pm$ 0.18 & 0.08 $\pm$  0.01\\
    HOPS 297      &  15.5 $\pm$ 1.7 & 1.00 $\pm$ 0.17 & 0.06 $\pm$  0.01\\
    HOPS 321      &  16.4 $\pm$ 1.7 & 0.54 $\pm$ 0.10 & 0.03 $\pm$  0.01\\
    HOPS 360      &  17.6 $\pm$ 1.8 & 1.52 $\pm$ 0.24 & 0.09 $\pm$  0.02
\enddata

\tablecomments{The column densities are calculated
	       assuming $T_{\rm ex}$ = 10 K.  The estimated errors
	       come from the uncertainties of integrated intensities
	       (Table \ref{table_obs_results}), which include the
	       statistical uncertainties and the calibration
	       uncertainty of 10\% \citep{Lee:2011dn}.}

\label{table_cd_ratio}
\end{deluxetable}

\subsection{HDCO column density}

\hdco\ is a nearly symmetric top molecule. Therefore, the treatment
for calculating column density is similar to the case of \htco.
However, there is no H-pair, so there is no ortho/para modification.
While \htco\ has electric dipole moment along the A axis only, HDCO
has an extra dipole moment along the B axis.
As a result, transitions between K ladders are allowed,
and all K ladders are radiatively connected.
\hdco\ lines have hyperfine structures \citep{Langer:1979fn}.  For
the \hdco\ \tranii\ transition, the central group (2 components)
dominates in intensity, and the velocity separation from center to
satellites is $< 0.2$ \kms. The hyperfine components are not expected
to be resolved, but they can make the line look wide. For the purpose
of column density calculations, we ignore the effects of hyperfine
structure. We use $\mu_{a}$ (2.324 Debye) for $\mu$, since we are
interested in the transitions in a K ladder (a type). The statistical
weight $g$ has a rotation contribution only, $g(J_{\rm K_{\rm
a},K_{\rm c}}) = 2J+1$. The HDCO transition is optically
thin because deuterated species are 10--100 times less
abundant than the main species.

The total HDCO column density is calculated by assuming an excitation
temperature of 10 K and optically thin line (Table \ref{table_cd_ratio}).
If we use $T_{\rm ex} =$ 30 K or 50 K, the HDCO column densities
change by a factor of about 3.4 or 6.7, respectively. The
[HDCO]/[H$_2$CO] ratios change by a factor of about 1.6 or 1.8,
respectively.

\subsection{Excitation temperature}
\label{sec:Tex}

\begin{deluxetable}{ccc}
\tabletypesize{\small}
\tablecaption{Excitation Temperatures}
\tablewidth{0pt}
\tablehead
{\colhead{Source} &
\colhead{$RT$\tablenotemark{a}} &
\colhead{$T_{\rm ex}$\tablenotemark{b}} \\
& & \colhead{(K)} }
\startdata
    119019     &1.34  & N                 \\
    HOPS 358   &0.21  & ~6 ($\pm$~1)      \\
    HOPS 373   &0.24  & ~7 ($\pm$~3)      \\
    093005     &0.58  & 31 ($\pm$18)      \\
    HOPS 359   &0.39  & 12 ($\pm$~4)      \\
    097002     &1.41  & N                 \\
    HOPS 354   &0.80  & N                 \\
\enddata

\tablenotetext{a}{The line ratio
                  of the HDCO \tranii\ and \traniii\ transitions.}
\tablenotetext{b}{N: $T_{\rm ex}$ cannot be determined because RT $>$ 3/4.
                  The qualities of their HDCO \traniii\ spectra
                  are relatively poor (Figure \ref{fig_spectra_pbrs}).}

\label{table_tex}
\end{deluxetable}

We calculated the excitation temperature
using the line ratio of the \hdco\ \jk = \tranii\ and \traniii\ transitions,
assuming that they have the same excitation temperature.
Because the $J$ quantum numbers are the same,
factors depending on $J$ are common factors
in calculating the column density ratio.
The column density ratio, ignoring the common factors, is
\begin{equation}
\frac{N_{1_{1,0}}}{N_{1_{0,1}}} = RT\frac{4}{3}
\frac{[1-{\rm exp}(-h\nu(2_{0,2}-1_{0,1})/kT_{\rm
ex})]}{[1-{\rm exp}(-h\nu(2_{1,1}-1_{1,0})/kT_{\rm ex})]} 
\approx RT\frac{4}{3},
\end{equation}
where $RT={\int_{}T_{\rm mb}(2_{1,1}-1_{1,0}){\rm d}v}/{\int_{}T_{\rm
mb}(2_{0,2}-1_{0,1}){\rm d}v}$ is the line ratio. Because the frequencies
are very similar, the right exponential factor after $RT4/3$ is
nearly 1.  The column density ratio should be the same as the ratio
of partition terms, and here the statistical weight $g$ cancel out
since $g$ only depends on $J$.
\begin{equation}
RT\frac{4}{3}=
{\rm exp}\left(-\frac{E(1_{1,0})-E(1_{0,1})}{kT_{\rm ex}}\right).
\end{equation}
Since $E(1_{1,0})/k = 11.18297$ K and $E(1_{0,1})/k = 3.09406$ K,
the excitation temperature is,
\begin{equation}
T_{\rm ex} = \frac{-8.09}{{\rm ln}(4/3 RT)}. 
\end{equation}

In our observations, seven sources are detected in both transitions.
When the line ratio is close to $\sim$0.75, $T_{\rm ex}$ diverges.
The excitation temperatures are summarized in Table \ref{table_tex}.
We obtained an average excitation temperature of $\sim$14 $\pm$ 5 K.
Therefore, the $T_{\rm ex} = 10$ K used in the column density
calculations is consistent with the measured $T_{\rm ex}$.

\subsection{Optical depths and line widths}
\label{sec:tau}


\begin{figure}[!t]
\epsscale{1}
\plotone{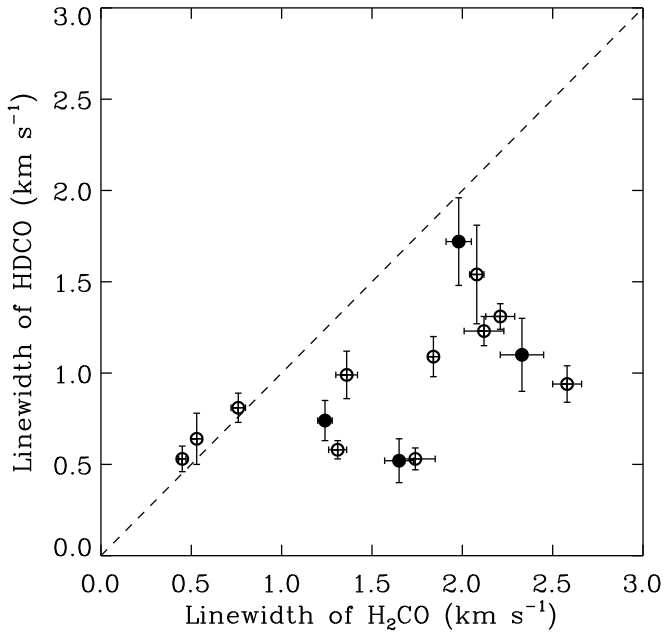}
\caption{Line widths of the H$_2$CO and HDCO spectra.
         Open and filled circles represent PBRS and non-PBRS, respectively.
         The dashed line represents equal widths.}

\label{fig_fwhm}
\end{figure}

The assumption of optically thin line is reasonable for the lines of the
deuterated molecules based on their low abundances. However, the
optical depth of the \htco\ \trani\ transition can be high because
\htco\ is abundant in molecular clouds. The optical depth of the
\htco\ line can be estimated by comparing the fluxes between
\htco\ and its isotopologue for the same transition.
Unfortunately, our survey did not cover observations of the
H$_2$$^{13}$CO line.  Though a direct measurement of the optical depth
($\tau$) is not available, we have roughly estimated the line optical
depths using the solution of the line radiative transfer equation.
\begin{equation}
\tau = -{\rm ln}(1 - (T_{\rm peak}/J_{\nu}(T_{\rm ex}) - J_{\nu}(T_{\rm BG}))), 
\end{equation}
where $T_{\rm peak}$ is the peak temperature, $J_{\nu}(T_{\rm ex})$
and $J_{\nu}(T_{\rm BG})$ are the equivalent Rayleigh-Jeans excitation
and background temperatures, respectively. The optical depths
estimated in this way are between 0.3 to 1.9. In other studies, the
optical depths of 0.4--1.1 were estimated from low-mass protostars
\citep[e.g.,][]{Maret:2004io} and $\sim$1.6 from the dense core in the
Horsehead PDR \citep{Guzman:2011da}, based on the flux ratio of the
\htco\ and H$_2$$^{13}$CO \trani\ transition. Larger line width
is also known to be coupled with larger line optical depth 
\citep{Phillips:1979ca, Roberts:2002ck, Ceccarelli:2003ir}.  In
Figure \ref{fig_fwhm}, we compare the line widths of \htco\ \trani\
and \hdco\ \tranii. Most sources with large \htco\ line widths do
not show a good agreement, which is likely due to the optical depth
in the \htco\ \trani\ transition. Considering the optical depth of the 
\htco\ estimated from our calculation, the column density of \htco\
should be corrected by a factor of $\tau/(1-{\rm exp}(-\tau))$,
then $N$(H$_2$CO) increases by a factor of 1.1 for $\tau = 0.3$ and
2.2 for $\tau = 1.9$.

\section{Discussion}

\subsection{Deuterium fractionation} 


\begin{figure}[!t]
\epsscale{1}
\plotone{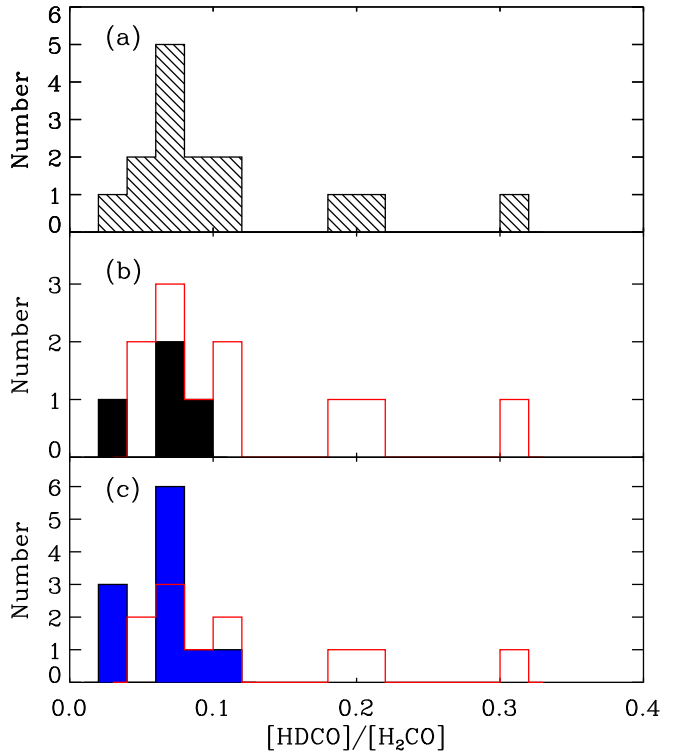}

\caption{Histograms of [HDCO]/[H$_2$CO].
(a) Distribution of the full sample.
(b) Distributions of PBRS (open bars) and non-PBRS (filled bars) samples.
(c) Distributions of PBRS (open bars) and non-PBRS (blue filled bars),
    including seven Class 0 sources from \cite{Roberts:2002ck}.}

\label{fig_DH_hist}
\end{figure}

The last column of Table \ref{table_cd_ratio} lists [HDCO]/[H$_2$CO]
ratios calculated assuming $T_{\rm ex}$ = 10 K. 
This ratio is not very sensitive to $T_{\rm ex}$ (Section 4.2). The [HDCO]/[H$_2$CO]
ratio ranges from 0.03 to 0.31, which is consistent with values
derived from the low-mass Class 0 protostars by \cite{Parise:2006bl}.
The average [HDCO]/[H$_2$CO] ratio and standard deviation of the total
target sample are 0.105 and 0.075, respectively. Figure
\ref{fig_DH_hist}(a) shows the distribution of the [HDCO]/[H$_2$CO]
ratios for the full sample.
Considering that the ratio is always positive
and that the distribution is asymmetric,
the distribution can be characterized better with the mode than the average.
The mode value and FWHM are $\sim$0.07 and $\sim$0.02, respectively.
The [HDCO]/[H$_2$CO] ratios of the three PBRS on the high end
(090003, HOPS 373, and 097002) are
significantly larger than the mode value.
They may represent either a long tail of a unimodal distribution
or a secondary peak of a bimodal distribution.
More extensive studies are necessary
to understand the detailed shape of the distribution.

Figure \ref{fig_DH_hist}(b) shows the [HDCO]/[H$_2$CO] distributions
for the PBRS and non-PBRS samples separately. About 30\% of
PBRS (3/11) have high [HDCO]/[H$_2$CO] ratios ($> 0.15$), and 70\%
of PBRS (8/11) have ratios similar to non-PBRS. Both PBRS and
non-PBRS have the same mode ($\sim$0.07) as the full sample.

Since the number of non-PBRS sources are relatively small,
the [HDCO]/[H$_2$CO] ratios of Class 0 sources
measured by \cite{Roberts:2002ck}
may be helpful in understanding the distributions.
Figure \ref{fig_DH_hist}(c) shows
the [HDCO]/[H$_2$CO] distribution of non-PBRS sources,
including 7 sources from \cite{Roberts:2002ck}.
The general properties of the distribution remain the same.
The mode is $\sim$0.07, and the range of values is less than $\sim$0.12,
which reinforces the finding
that the PBRS distribution has a relatively large dispersion
and that the three PBRS on the high end are outliers.
However, note that five sources in the sample of \cite{Roberts:2002ck}
are in regions closer than the Orion cloud,
which could introduce some unexpected effects to the distribution.
Therefore, a more extensive survey with a larger sample size
will be helpful in understanding the distributions better.


\begin{figure}[!t]
\epsscale{1}
\plotone{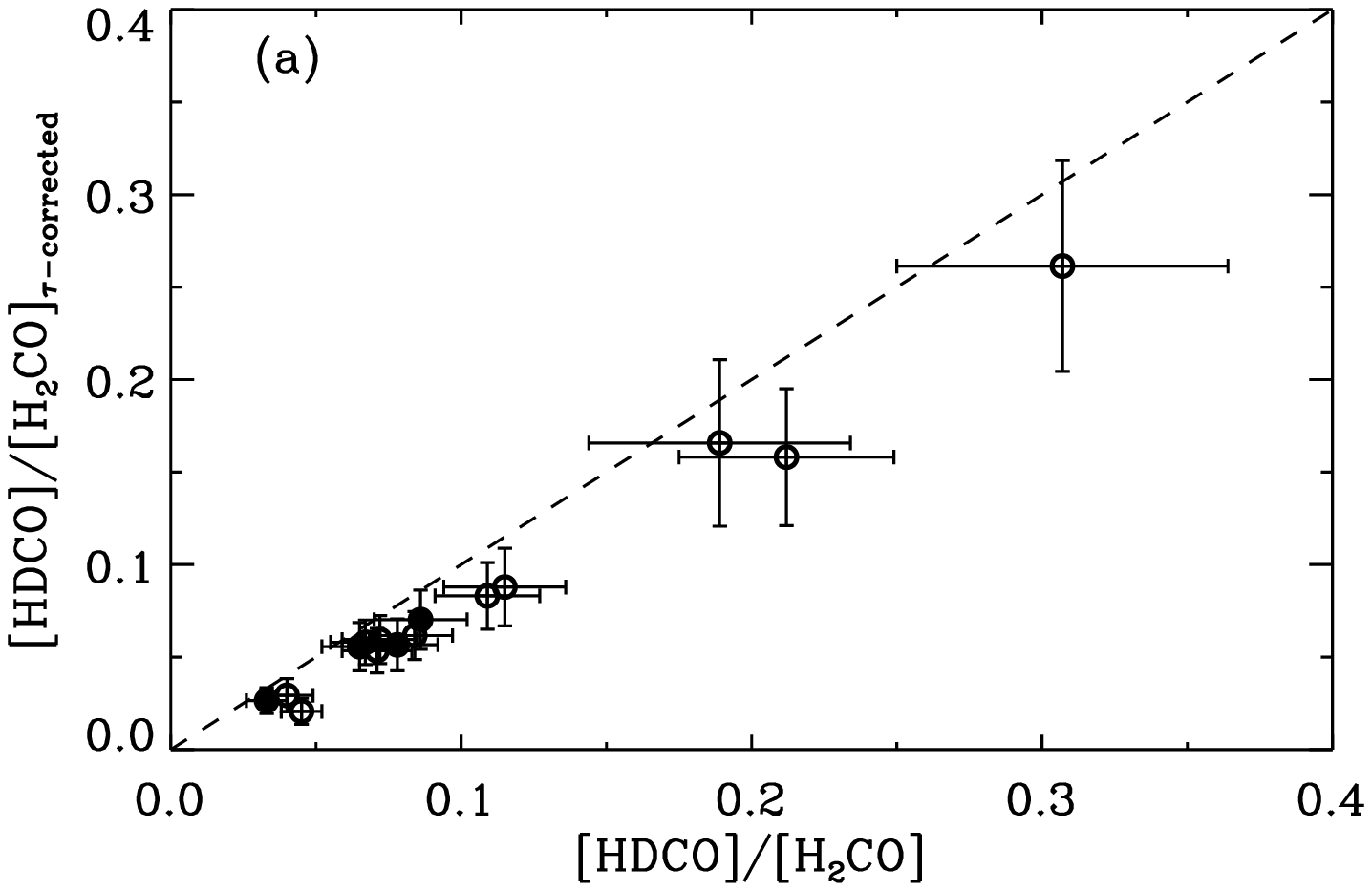}
\plotone{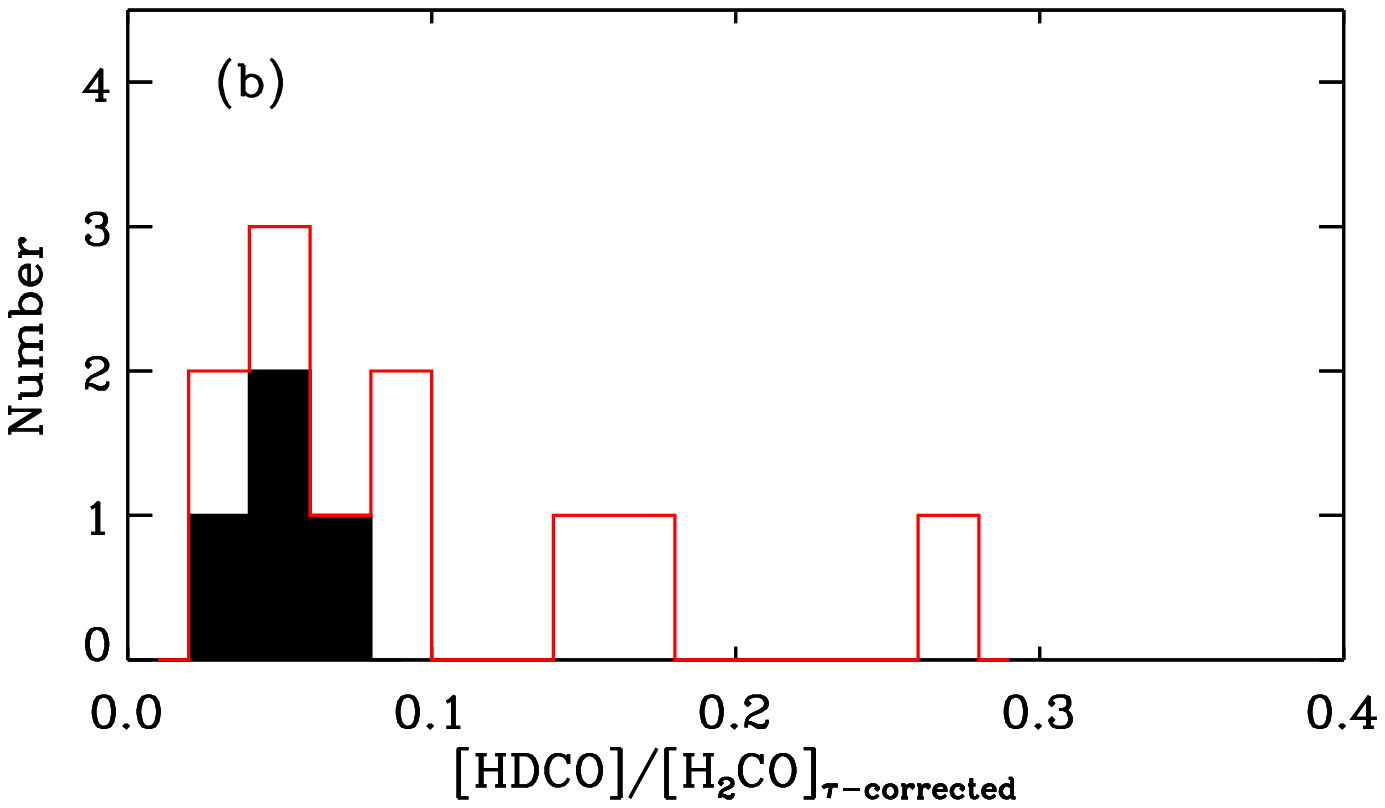}

\caption{(a) Comparison of uncorrected [HDCO]/[H$_2$CO] ratios with
$\tau$-corrected [HDCO]/[H$_2$CO] ratios. Open and filled circles
represent PBRS and non-PBRS, respectively. The dashed line represents
equal [HDCO]/[H$_2$CO] ratio. (b) Histograms of $\tau$-corrected
[HDCO]/[H$_2$CO].}

\label{fig_tau}
\end{figure}

Figure \ref{fig_tau}(a) shows a comparison of the [HDCO]/[H$_2$CO] ratios
with and without optical depth correction.
Although the optical depth corrections for the H$_2$CO
observations make the [HDCO]/[H$_2$CO] ratios decrease for all
sources, changes of the [HDCO]/[H$_2$CO] ratios are not much. In
Figure \ref{fig_tau}(b), the overall distributions of the
$\tau$-corrected [HDCO]/[H$_2$CO] ratios show similar characteristics
compared with Figure \ref{fig_DH_hist}(b). The three outliers with
high deuterium fractionation are still far from the mode of the
distribution. Even if we consider the optical depth, the [HDCO]/[H$_2$CO]
ratios of 090003, HOPS 373, and 097002 are still high ($>$ 0.15).
Therefore, in the subsequent discussions below, we will consider
only the [HDCO]/[H$_2$CO] distribution derived with the assumption
of optically thin lines.

\begin{figure*}[!t]
\epsscale{1.5}
\plotone{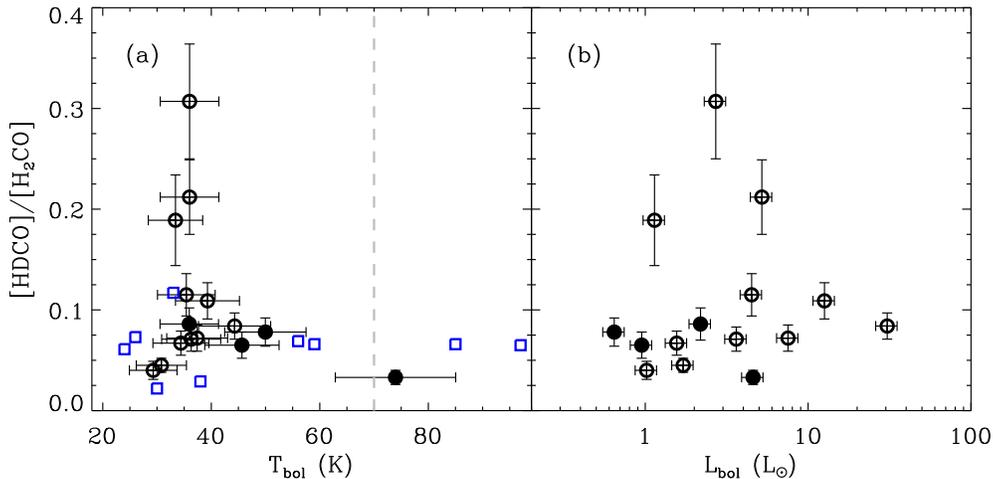}

\caption{
[HDCO]/[H$_2$CO] ratios of PBRS (open circle) and non-PBRS (filled circle)
plotted against (a) the bolometric temperature and (b) the luminosity.
Blue square symbols represent the low-mass protostellar cores
of \cite{Roberts:2002ck}.
The dashed line shows the canonical $T_{\rm bol}$ division
between the Class 0 and Class I sources \citep{Chen:1995eo}.}

\label{fig_dh_tbol}
\end{figure*}

The [HDCO]/[H$_2$CO] ratios are plotted against $T_{\rm bol}$
and $L_{\rm bol}$ in Figure \ref{fig_dh_tbol}.
When plotted against $L_{\rm bol}$ in Figure \ref{fig_dh_tbol}(b),
no obvious correlation was found.
The [HDCO]/[H$_2$CO]
ratios for low-mass Class 0 protostars reported by \cite{Roberts:2002ck}
are also plotted in Figure \ref{fig_dh_tbol}(a). As suggested by
\cite{Roberts:2002ck}, there is no marked correlation between the
[HDCO]/[H$_2$CO] ratios and the bolometric temperature
(a proxy of the evolutionary stage of protostars).
Though the PBRS and non-PBRS samples show similar mode values,
the PBRS sample shows a significantly larger scatter.
This difference suggests
that either PBRS have an intrinsically large variation
in the deuterium fractionation
or the three outliers are chemically distinct
from the majority of Class 0 protostars.


\subsection{Properties of the PBRS sample}

To focus on the large variation of the deuterium fractionation in
the PBRS sample, we calculate the fractional abundances of the PBRS
sample. The fractional HDCO and \htco\ abundances were calculated
by dividing the total HDCO and \htco\ column densities by the H$_2$
column density i.e., X(HDCO) = [HDCO]/[H$_2$] and X(H$_2$CO) =
[H$_2$CO]/[H$_2$].
Using the 870 \micron\ flux density in Table 5 of \cite{Stutz:2013ka},
the beam-averaged column density of PBRS can be computed with:
\begin{equation}
N_{\rm H_2} = \frac{F_\nu R}{B_\nu(T_D)\Omega \kappa_\nu \mu m_{\rm H}},
\end{equation}
where $\Omega$ is the beam solid angle, $\mu$ is the mean molecular
weight of the interstellar medium, which we assume to be 2.8
\citep{Kauffmann:2008jj}, $m_{\rm H}$ is the mass of the hydrogen
atom, $\kappa_\nu$ is the dust absorption coefficient (1.85
cm$^2$g$^{-1}$, \cite{Schuller:2009ef}), and $R$ is the gas-to-dust
mass ratio (assumed to be 100). We assume a dust temperature of 20
K \citep{Stutz:2010kc}. The observed abundances of HDCO and H$_2$CO
range from $2\times10^{-11}$ to $7\times10^{-11}$ and $8\times10^{-11}$
to $1\times10^{-9}$, respectively. The H$_2$CO abundances of PBRS
are similar to the outer H$_2$CO abundances toward low-mass protostars
reported by \cite{Maret:2004io}.  They have explained that the outer
regions of the envelopes of Class 0 sources reflect the pre-collapse
conditions based on the similarity of the values between the
abundances of pre-stellar cores studied by \cite{Bacmann:2002db,
Bacmann:2003kv} and the outer \htco\ abundances of low-mass protostars.


\begin{figure}[!t]
\epsscale{1.}
\plotone{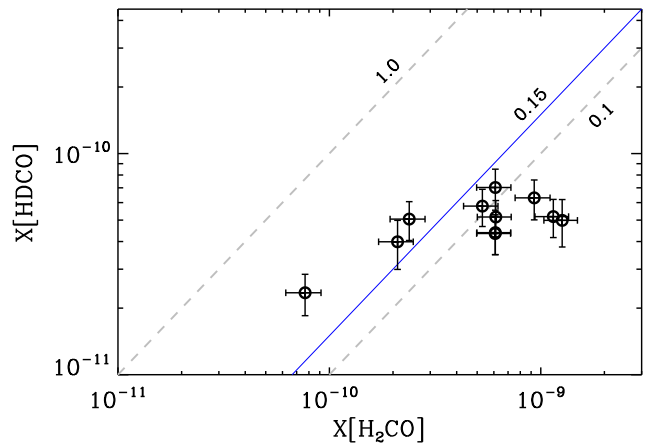}

\caption{Abundance ratios of \htco\ and HDCO relative to H$_2$
molecules for the PBRS sample. The gray dashed lines represent the
ratios of 1.0 and 0.10 from top to bottom. Three highly deuterated
PBRS are located on the left side of the blue solid line representing
the ratio of 0.15.}

\label{fig_xh2co_xhdco}
\end{figure}


\begin{figure}[!t]
\epsscale{1.}
\plotone{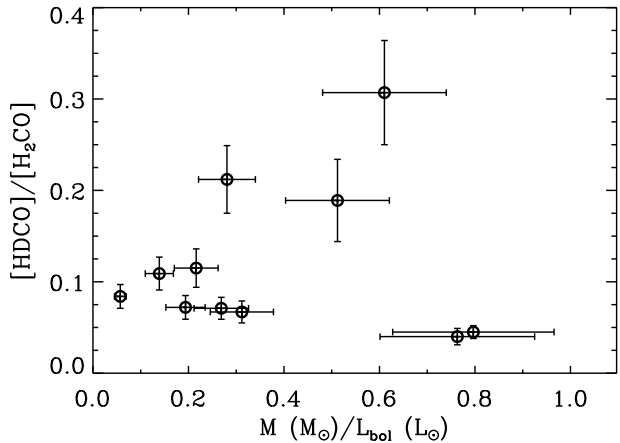}

\caption{
[HDCO]/[\htco] ratios plotted against the mass-to-luminosity ratios
for the PBRS sample.}

\label{fig_masslum_dh}
\end{figure}

The HDCO and H$_2$CO abundances relative to H$_2$ are plotted in
Figure \ref{fig_xh2co_xhdco}. As the \htco\ abundance increases,
the HDCO abundance increases for the PBRS with the ratio of $> 0.1$.
The highly deuterated PBRS ([HDCO]/[\htco] $>$ 0.15)
are significantly different from the other PBRS
in that they have small \htco\ abundances
and that their \htco\ line widths are quite narrow
($<$ 1 \kms, Figure \ref{fig_fwhm}).
The degree of deuterium fractionation is suggested to increase
with increasing CO depletion \citep{Bacmann:2003kv}.
The CO and \htco\ molecules are depleted
by a similar factor in the outer envelope of Class 0 protostar
\citep{Maret:2004io}. The low \htco\ abundance of highly deuterated
PBRS is likely to be explained by depletion of \htco.
The narrow line widths imply that these cores are quiescent,
which is associated with early phases of core evolution.
Therefore, the PBRS with high [HDCO]/[\htco] ratios
are probably in the very earliest stage of star formation.

Figure \ref{fig_masslum_dh} shows the relation
between [HDCO]/[\htco] and the mass-to-luminosity ratio
that is expected to decrease with time.
The PBRS with high [HDCO]/[\htco] ratios
have relatively large mass-to-luminosity ratio,
supporting that they are in the earliest stage of star formation.
However, two PBRS with the largest mass-to-luminosity ratio
have low [HDCO]/[\htco] ratios.
Therefore, it is difficult to find a simple relation
between the two quantities.
If the mass-to-luminosity ratio is a good tracer of evolution,
[HDCO]/[\htco] seems to have a large dispersion at early stages
and converge to a typical value of Class 0 sources at later stages.

The large variation of deuterium fractionation in the whole PBRS
sample leads us to suggest that the PBRS form in diverse conditions
or have diverse formation histories.  Indeed, the degree of deuterium
fractionation of \htco\ is sensitive to the initial D/H ratios of
gaseous molecules determined before the collapse phase
\citep{Turner:1990dg, Charnley:1992ku, Roberts:2002ck, Cazaux:2011dt,
Aikawa:2012di}.

PBRS 097002 and 090003 among the highly deuterated PBRS have flat
visibility amplitude profiles in the 2.9 mm dust continuum observations
by \cite{Tobin:2015jv}. They interpret that the PBRS with flat
visibility amplitude profiles are the youngest in a brief phase of
high infall/accretion. Since PBRS 093005 and 082005 with the lowest
[HDCO]/[H$_2$CO] ratio also have flat visibility amplitude profiles,
it is unclear if the variation of deuterium fractionation in \htco\
is due to the core evolution. Additional observations in pure gas
phase species, e.g., N$_2$D$^+$ and N$_2$H$^+$, can help understanding
the chemical evolutionary sequence of PBRS.

\section{Summary}

We observed 15 Class 0 protostars (11 PBRS and 4 non-PBRS) in
the Orion molecular cloud complex using the KVN in the single-dish
telescope mode. The chemical properties of PBRS were investigated
in the \htco\ \trani, \hdco\ \tranii, and \tranii\ lines. The main
results are summarized as follows.

\begin{enumerate}

\item The \htco\ \trani\ and \hdco\ \tranii\ emission lines were detected
toward all targets. The \hdco\ \traniii\ emission line was detected toward
seven PBRS.

\item The [HDCO]/[H$_2$CO] ratio ranges from 0.03 to 0.31. The
deuterium fractionation of most PBRS (70\%) is similar to that of
non-PBRS Class 0 protostars, and three PBRS (30\%) have significantly
high deuterium fractionation, greater than 0.15.
These findings represent a good hint
that the deuterium fractionation of PBRS
may have a distribution different from that of non-PBRS,
but future studies with larger sample sizes are needed
to reach a statistically more significant conclusion.

\item The high [HDCO]/[H$_2$CO] ratios of the three PBRS imply
that they are in the earliest phase of the star formation.
The large variation of deuterium fractionation
in the whole PBRS sample suggests
that PBRS form in diverse conditions of temperature and/or density,
because [HDCO]/[H$_2$CO] ratios are usually interpreted
as a fossil memory of the earlier, colder phase of star formation.

\item No clear correlation between the deuterium fractionation of
\htco\ and the evolutionary state of cores could be found. Further
studies in various molecules are needed to investigate the difference
between evolutionary sequence and initial conditions of the PBRS
sample.

\end{enumerate}

\acknowledgments

We thank Friedrich Wyrowski, Thomas Megeath, B\'{e}reng\`{e}re Parise,
Jeong-Eun Lee, and Karl M. Menten for helpful discussions.


\end{document}